\documentclass[conference]{IEEEtran}

\usepackage{cite}

\ifCLASSINFOpdf
   \usepackage[pdftex]{graphicx}
\else
   \usepackage[dvips]{graphicx}
\fi

\usepackage{float}
\usepackage{siunitx}
\usepackage{booktabs}
\usepackage{makecell}
\usepackage{multicol}

\usepackage{array}

\usepackage[tight,footnotesize]{subfigure}

\usepackage[font=footnotesize]{subfig}

\usepackage{midfloat}
\usepackage{caption}

\usepackage{stfloats}

\usepackage{cleveref}

\begin{document}
%
\title{On the Resilience of Underwater\\Semantic Wireless Communications\vspace{-1em}}

\author{\IEEEauthorblockN{João~Pedro~Loureiro\IEEEauthorrefmark{1},
        Patrícia~Delgado\IEEEauthorrefmark{1},
        Tomás~Feliciano~Ribeiro\IEEEauthorrefmark{2},
        Filipe~B.~Teixeira\IEEEauthorrefmark{1},
        Rui~Campos\IEEEauthorrefmark{1}
        \\
        }
        \IEEEauthorblockA{
        \IEEEauthorrefmark{2}\IEEEauthorrefmark{1}INESC TEC and Faculdade de Engenharia, Universidade do Porto\\
Campus da FEUP, Rua Dr. Roberto Frias, s/n - 4200-465 Porto, Portugal\\
\IEEEauthorrefmark{1}\{joao.p.loureiro, patricia.v.delgado, filipe.b.teixeira, rui.l.campos\}@inesctec.pt, \IEEEauthorrefmark{2}\{up202207427\}@edu.fe.up.pt
\thanks{This work is financed by National Funds through the Portuguese funding agency, FCT - Fundação para a Ciência e a Tecnologia, within project ACOUSTNET (LA/P/0063/2020) and the scholarship 2022.14283.BD.}}
}

\IEEEoverridecommandlockouts
\maketitle

\IEEEpeerreviewmaketitle

\section*{Abstract}
Underwater wireless communications face significant challenges due to propagation constraints, limiting the effectiveness of traditional radio and optical technologies. Long-range acoustic communications support distances up to a few kilometers, but suffer from low bandwidth, high error ratios, and multipath interference. Semantic communications, which focus on transmitting extracted semantic features rather than raw data, present a promising solution by significantly reducing the volume of data transmitted over the wireless link.

This paper evaluates the resilience of SAGE, a semantic-oriented communications framework that combines semantic processing with Generative Artificial Intelligence (GenAI) to compress and transmit image data as textual descriptions over acoustic links. To assess robustness, we use a custom-tailored simulator that introduces character errors observed in underwater acoustic channels. Evaluation results show that SAGE can successfully reconstruct meaningful image content even under varying error conditions, highlighting its potential for robust and efficient underwater wireless communication in harsh environments.

\section{Introduction}
Underwater wireless communications are fundamentally constrained by harsh propagation conditions \cite{survey,UW_Filipe,teixeira2021novel}. Yet, reliable underwater communication systems remain vital for advancing the Blue Economy and enabling essential applications, including environmental monitoring, oceanographic exploration, and marine robotics \cite{EconomiaAzul,DURIUS,oceans21}. Short-range communications such as radio frequency (RF) and wireless optical communications can offer high data rates, but their performance rapidly degrades with distance due to the medium's properties \cite{survey}. For long-range, acoustic communications remain the preferred solution, providing underwater connectivity over distances of several kilometers \cite{acoustic_modems}. Nonetheless, acoustic channels still face severe limitations, including low bandwidth, high latency, and high transmission error ratios.

Semantic communications have emerged as a promising approach to mitigate bandwidth and reliability constraints in wireless networks \cite{DBLP:journals/corr/abs-2201-01389}. Instead of transmitting raw data (e.g., images), semantic systems encode and transmit high-level features that capture the underlying meaning of the information. This significantly reduces the data volume, making semantic communication particularly suitable for error-prone, low-throughput environments such as underwater acoustic channels. Despite its potential, semantic communications in the underwater environment remain largely unexplored. Only a few preliminary works have investigated its feasibility  \cite{Xu:24,10248874,IMTS_Semm_Comms_2025}, and its performance under realistic channel impairments remains an open research challenge.

\begin{figure}[t]
\includegraphics[width=0.48\textwidth]{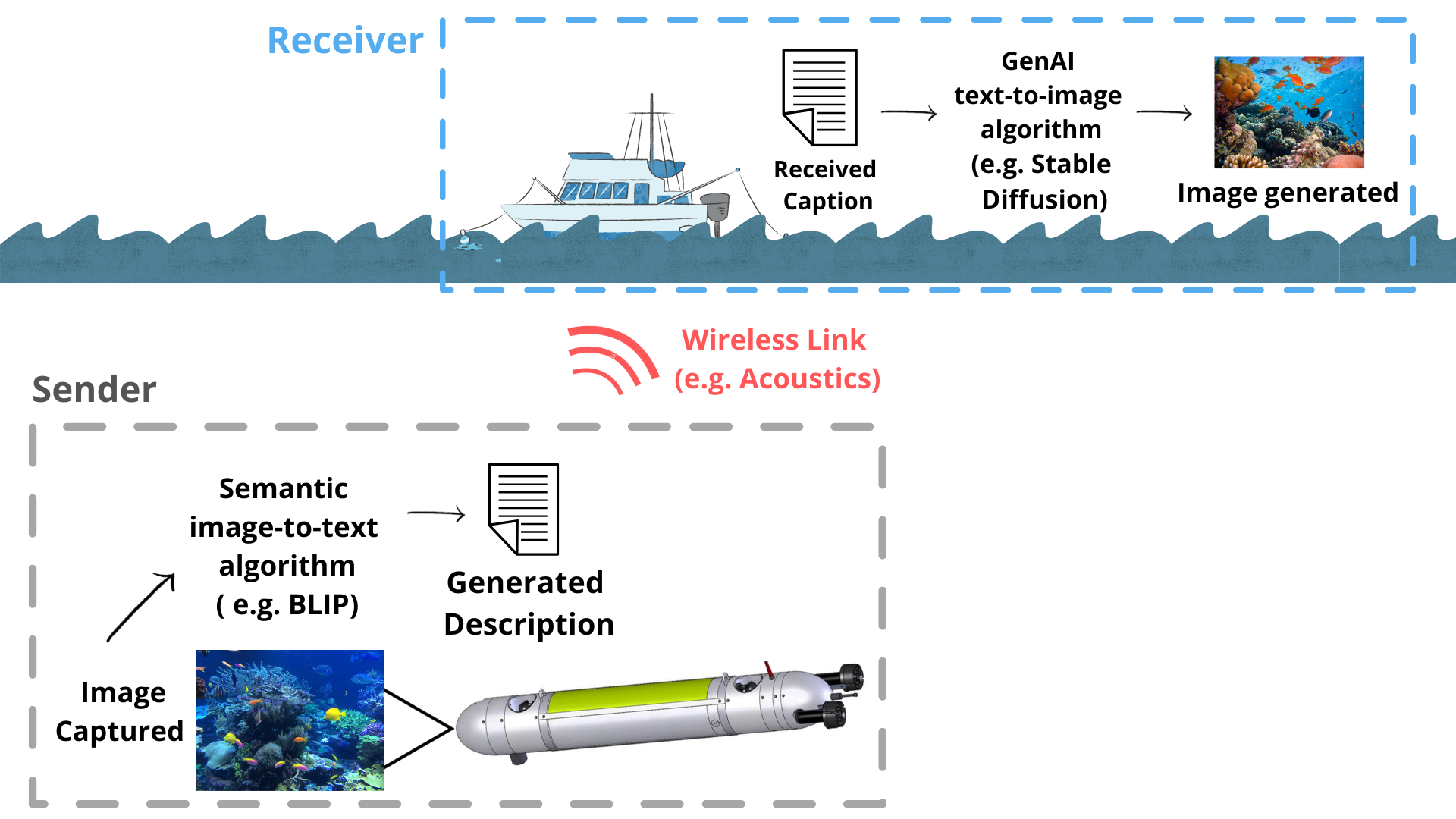}
\centering
\caption{High-level diagram of the SAGE framework for underwater semantic wireless communications \cite{sage}.}
\captionsetup{justification=centering}  
\label{fig:1}
\end{figure}

In our previous work, we introduced SAGE, a semantic communication framework tailored for underwater scenarios \cite{sage}. SAGE leverages image-to-text and text-to-image generative models to encode visual content into compact textual descriptions. At the transmitter, a GenAI model generates a description from an image, which is then transmitted acoustically. At the receiver, another GenAI model reconstructs the image based on the received description (see Fig.~\ref{fig:1}). This process enables a huge compression from a typical 2–10 MB image size to the size of a Short Message Service (SMS) message, in the order of 100 bytes. SAGE is model-agnostic, allowing flexible integration of state-of-the-art generative models and potential extension to other media types, such as video. 

In this paper, we evaluate the resilience of SAGE under noisy underwater acoustic conditions. Specifically, we assess the performance of two GenAI models -- BLIP for image-to-text conversion and Stable Diffusion for text-to-image reconstruction -- when exposed to common text-level transmission errors. These include random character errors, character deletions, and word deletions. To simulate realistic transmission conditions, we developed a simple configurable error injection simulator. The evaluation results show that SAGE can reconstruct semantically meaningful image content, even under significant error conditions, by leveraging GenAI’s ability to infer missing or corrupted information.

\vspace{10pt}
The major contributions of this work are:
\begin{enumerate}
    \item \textbf{Resilience Analysis of SAGE} -- We analyze the resilience of the SAGE framework under various types and levels of transmission errors;
    \item \textbf{Performance Evaluation of GenAI Models} -- We evaluate the performance of BLIP and Stable Diffusion in reconstructing visual content from degraded semantic descriptions;
    \item \textbf{Demonstration of Semantic Communication Robustness} -- We provide empirical evidence that semantics can maintain robust performance in challenging underwater environments, supporting its potential for future underwater wireless systems.
\end{enumerate}

The rest of the paper is organized as follows. Section II revises the state-of-the-art in underwater semantic communications and summarizes relevant models proposed in the literature. Section III describes the custom-tailored channel error simulator and presents the evaluation results. Finally, Section IV draws the main conclusions and points out future work.

\section{State of the Art}
\begin{figure}[t]
\includegraphics[width=0.48\textwidth]{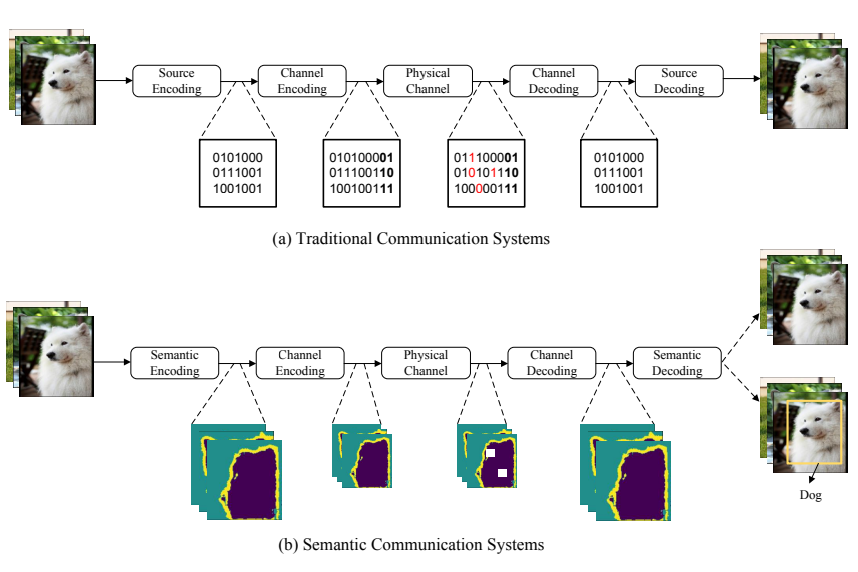}
\centering
\caption{Comparison between a semantic communications system (a) and a conventional one (b).\cite{DBLP:journals/corr/abs-2201-01389}.}
\captionsetup{justification=centering}  
\label{fig:2}
\end{figure}

This work extends our previous publication, in which we introduced SAGE -- a semantic communication framework tailored for underwater environments \cite{sage}. SAGE leverages semantic processing and GenAI to enable image transmission over low-bandwidth acoustic channels by encoding and transmitting only compact textual descriptions. This approach signinficantly reduces payload size compared to traditional source coding methods. Instead of transmitting raw or compressed image data, it transmits semantically rich text, which is then used to reconstruct the image at the receiver using GenAI models. Semantic communications represent a paradigm shift in wireless systems by focusing on transmitting the meaning of data, rather than the data itself. This reduces transmission throughput while preserving the core intent or information \cite{DBLP:journals/corr/abs-2201-01389}. As illustrated in Fig. \ref{fig:2}, semantic systems differ from conventional architectures by incorporating intelligence at the transmitter and receiver to extract and interpret high-level features. 

Besides SAGE, semantic communication approaches for underwater wireless communications are still scarce in the literature. In \cite{10248874}, the authors introduce a semantic communications framework that aims to improve both efficiency and reliability in acoustic links, supported by a basic simulation model. Similarly, \cite{Xu:24} presents a semantic system for underwater wireless optical communication using deep learning to extract and transmit essential image features, achieving better performance than traditional approaches. More recently, in \cite{IMTS_Semm_Comms_2025}, a new acoustic semantic transmission approach is proposed. In this solution, image data is first converted into descriptive text, containing a list of key visual elements (e.g., objects and environmental features). This list, derived by removing superfluous words, such as pronouns, further reduces the transmitted message size at the cost of some descriptive detail. While these solutions provide valuable insights, they often overlook crucial aspects that could influence the accuracy and reliability of similarity evaluations. A deeper analysis is needed to fully understand the nuances of the different metrics.  

The core components of semantic systems often rely on Image-to-Text and Text-to-Image models. Among Image-to-Text solutions, Contrastive Language-Image Pretraining (CLIP)~\cite{DBLP:journals/corr/abs-2103-00020} performs zero-shot captioning by identifying semantic attributes of unseen visual content (e.g., objects, animals, people). ViLBERT~\cite{DBLP:journals/corr/abs-1908-02265} integrates image and text modalities through two parallel Bidirectional Encoder Representations from Transformers (BERT) streams, one dedicated to images and the other to text, allowing joint understanding of visual and linguistic elements. BLIP (Bootstrapping Language-Image Pre-training)~\cite{DBLP:journals/corr/abs-2201-12086} refines this approach by combining a captioning module with a filtering mechanism to generate high-quality synthetic image descriptions. Its successor, BLIP-2 \cite{BLIP2}, adopts a modular architecture that connects frozen image encoders with Large Language Models (LLMs), improving performance and scalability while reducing computational overhead.

Text-to-Image can be performed using diffusion models, Vision-Language Models (VLMs), or transformer-based architectures. DALL-E~\cite{DBLP:journals/corr/abs-2102-12092} models text and image tokens in a unified sequence, using a two-stage process: first, a visual codebook compresses the image into tokens; then a transformer generates images conditioned on textual input. Imagen \cite{saharia2022photorealistictexttoimagediffusionmodels} creates low-resolution images from text and enhances them through cascaded super-resolution steps. Stable Diffusion~\cite{DBLP:journals/corr/abs-2112-10752} is a prominent open-source model that uses a denoising diffusion process to generate high-fidelity images from textual prompts. Its open nature fosters community contributions, making it valuable for both research and commercial use.

\section{Evaluation of SAGE under Transmission Errors}

\begin{figure}[t]
\includegraphics[width=0.48\textwidth]{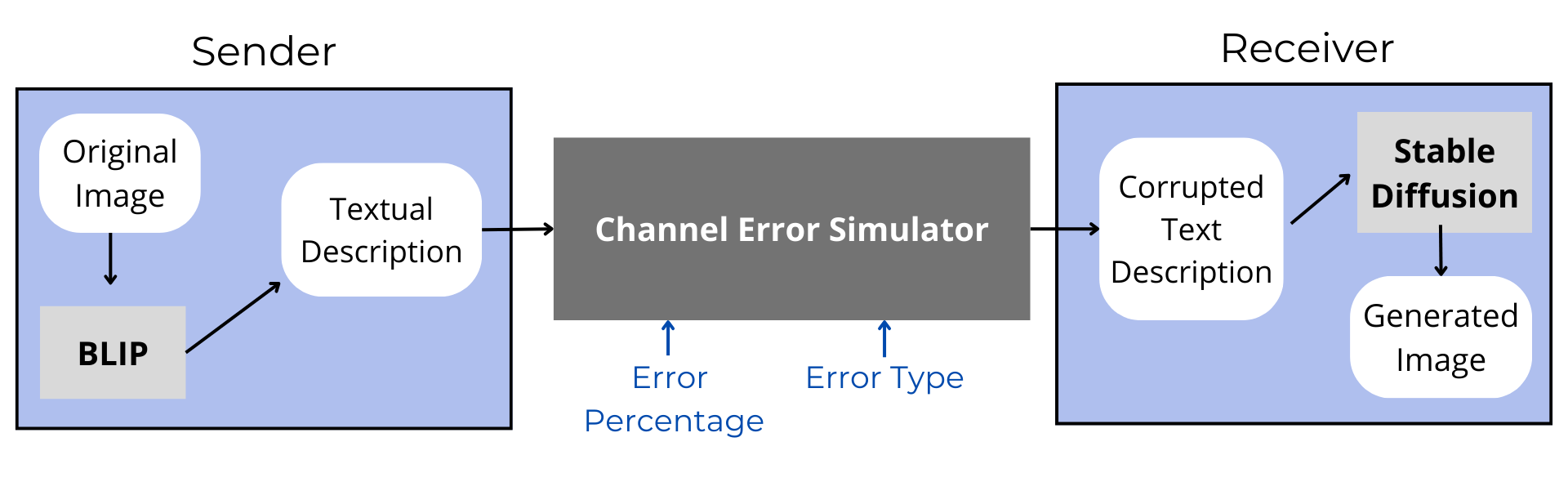}
\centering
\caption{Semantic communication pipeline employed for a single image transmission, used as a basis for the evaluation of SAGE under transmission errors.}
\captionsetup{justification=centering}  
\label{Testset_diag}
\end{figure}

To evaluate the resilience of semantic communications in underwater acoustic environments, we employed the SAGE framework alongside a custom-tailored text-level channel error simulator. An overview of the evaluation process is illustrated in Fig. \ref{Testset_diag}. 
Our test scenario involved transmitting a single image per test instance through the semantic pipeline. The dataset used for evaluation comprised 30 underwater images, the majority from the public dataset available in~\cite{10129222}. Each image was processed once using the BLIP model for image-to-text conversion. We avoided repeated runs, as multiple executions consistently yielded nearly identical captions. Conversely, since Stable Diffusion is inherently stochastic, we generated each image 10 times for the same description, both under error-free and corrupted conditions, to account for randomness. In total, $10 \times 30 = 300$ images were generated for each error ratio considered. 

\subsection{Text-level Channel Error Simulator}
\begin{figure}[b]
\includegraphics[width=0.48\textwidth]{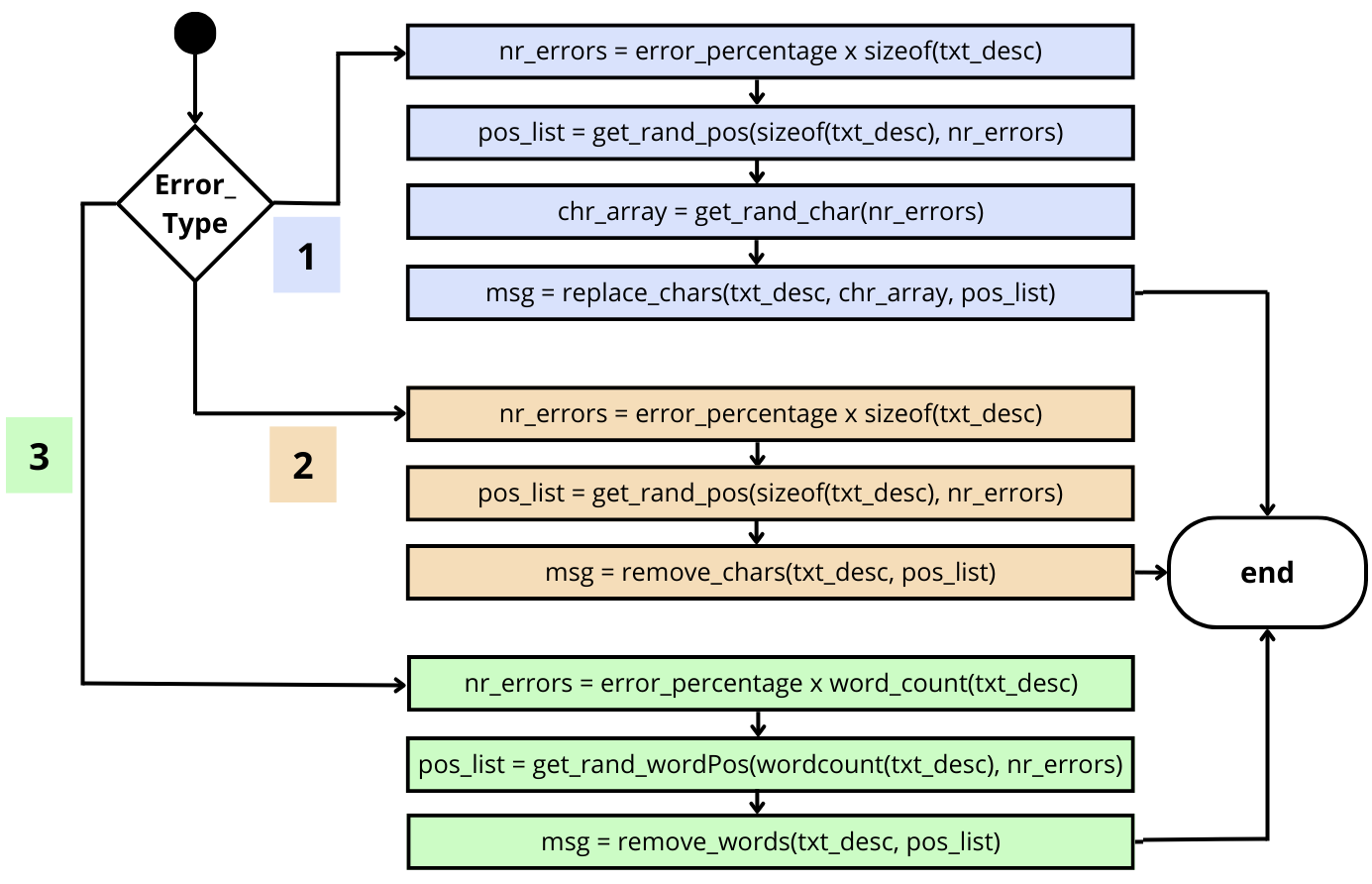}
\centering
\caption{Flowchart of the text-level channel error simulator used to introduce controlled transmission errors.}
\captionsetup{justification=centering}  
\label{simulator_diag}
\end{figure}

To simulate text-level transmission errors, we developed a simple simulator that injects textual errors before decoding the message at the receiver (see Fig.~\ref{simulator_diag}). The simulator supports the following error types:

\begin{enumerate}
    \item \textbf{Random Character Substitution} -- Random characters replace existing ones in the text;
    \item \textbf{Character Deletion} -- Random characters are deleted from the text;
    \item \textbf{Words Deleted} -- Entire words are deleted from the text.
\end{enumerate}

Each error type is simulated independently. The error ratio can be adjusted by an input variable, which defines the percentage of characters (types 1 and 2) or words (type 3) affected.

From a communications theory perspective, the error ratio corresponds to a range of possible BER, depending on how many bits are affected within each corrupted character. Assuming character consists of $b=8$ bits (i.e., one byte), the BER is bounded as follows:  
\[
BER \in [CER/b, CER]
\]  

where CER is the Character Error Ratio. The lower bound assumes that only one bit per corrupted character is affected, while the upper bound corresponds to the case where all bits in each corrupted character are erroneous. This means that depending on the error distribution, the effective BER for a 50\% CER, for instance, can range from 6.25\% (best-case scenario) to 50\% (worst-case scenario). These bounds help contextualize how text-level errors introduced by our simulator relate to conventional BER in communication systems.

\subsection{Evaluation Metrics}
To measure the impact of transmission errors on image reconstruction, we employed three similarity metrics:
\begin{itemize}
    \item Peak Signal-to-Noise Ratio (PSNR) \cite{PSNR};
    \item Structural Similarity Index (SSIM) \cite{SSIM};
    \item CLIP similarity score (CLIPScore) \cite{CLIP_similarity}.
\end{itemize}

While PSNR and SSIM assess pixel-level fidelity between the generated and original images, CLIPScore captures semantic similarity using multimodal embeddings. This distinction is critical when evaluating perceptual closeness versus literal image similarity. PSNR may range between 20 and 80 dB and SSIM values vary from 0 to 1, where higher values correspond to higher similarity in both cases. CLIPScore can range from 0\% to 100\%, representing similarity percentage (i.e., the higher percentage, the higher similarity).

\begin{figure*}
\includegraphics[width=0.9\textwidth]{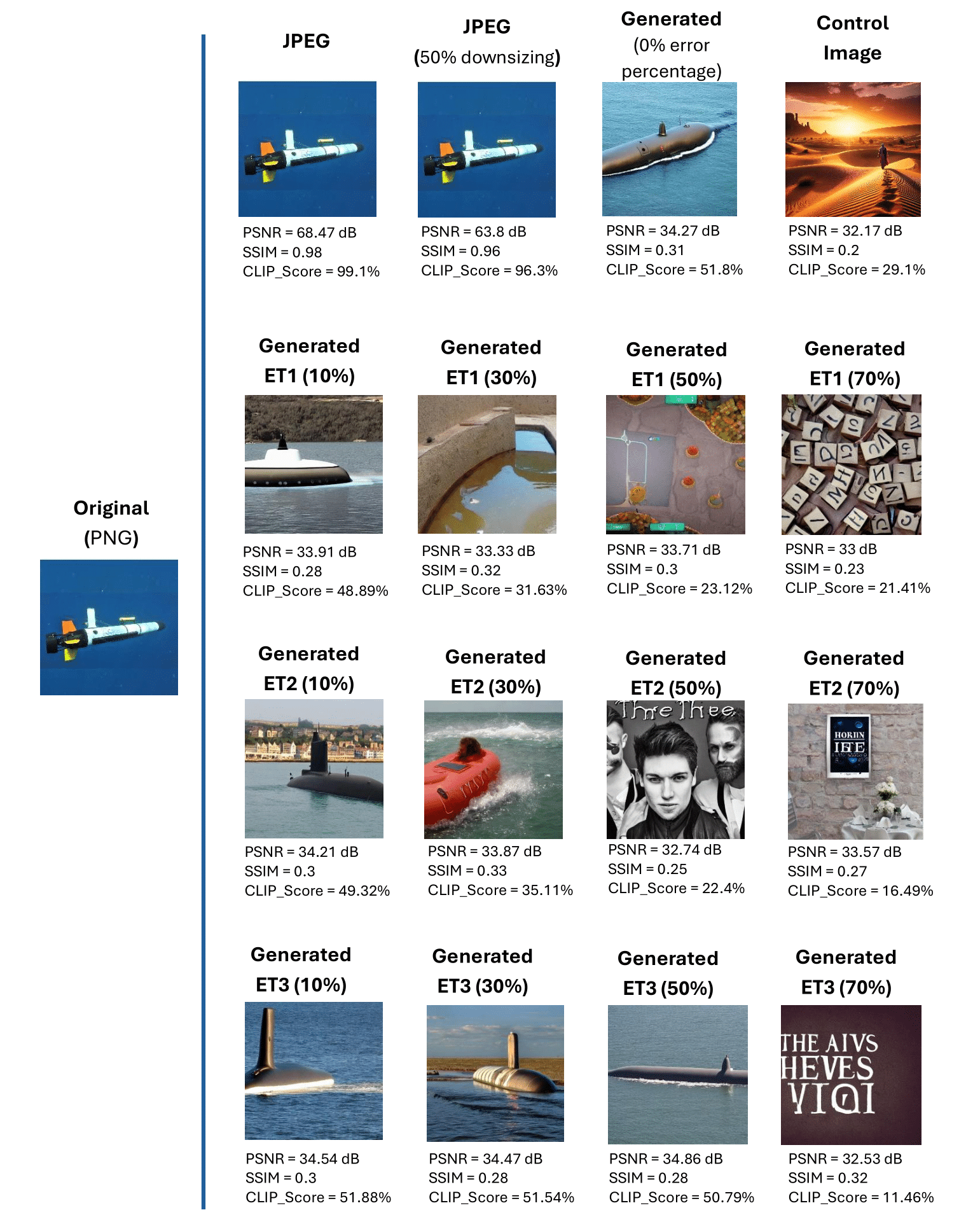}
\centering
\caption{Comparison of similarity metrics for the original image and: JPEG Compressed versions; a control image; and AI-generated images from text descriptions affected by different error types and ratios.}
\captionsetup{justification=centering}  
\label{tabela_exemplos}
\end{figure*}

We assessed the evolution of these metrics across increasing CER for each error type. For each test case, we evaluated:

\begin{itemize}
    \item Similarity between the original image and the generated image (blue line in plots);
    \item Similarity between a fixed control image and the generated image (red line).
\end{itemize}

The control image is a generic, semantically unrelated AI-generated image, used as a reference baseline to help interpret metric limits. The plots show the average metric score (y-axis) against the error percentage (x-axis) for a given error type. Fig.~\ref{tabela_exemplos} presents representative similarity scores alongside sample image outputs to illustrate these effects. These samples illustrate how different metrics can yield varying assessments of similarity, emphasizing the importance of considering multiple alternatives when evaluating image similarity.

\begin{figure}[h!]
\includegraphics[width=0.47\textwidth]{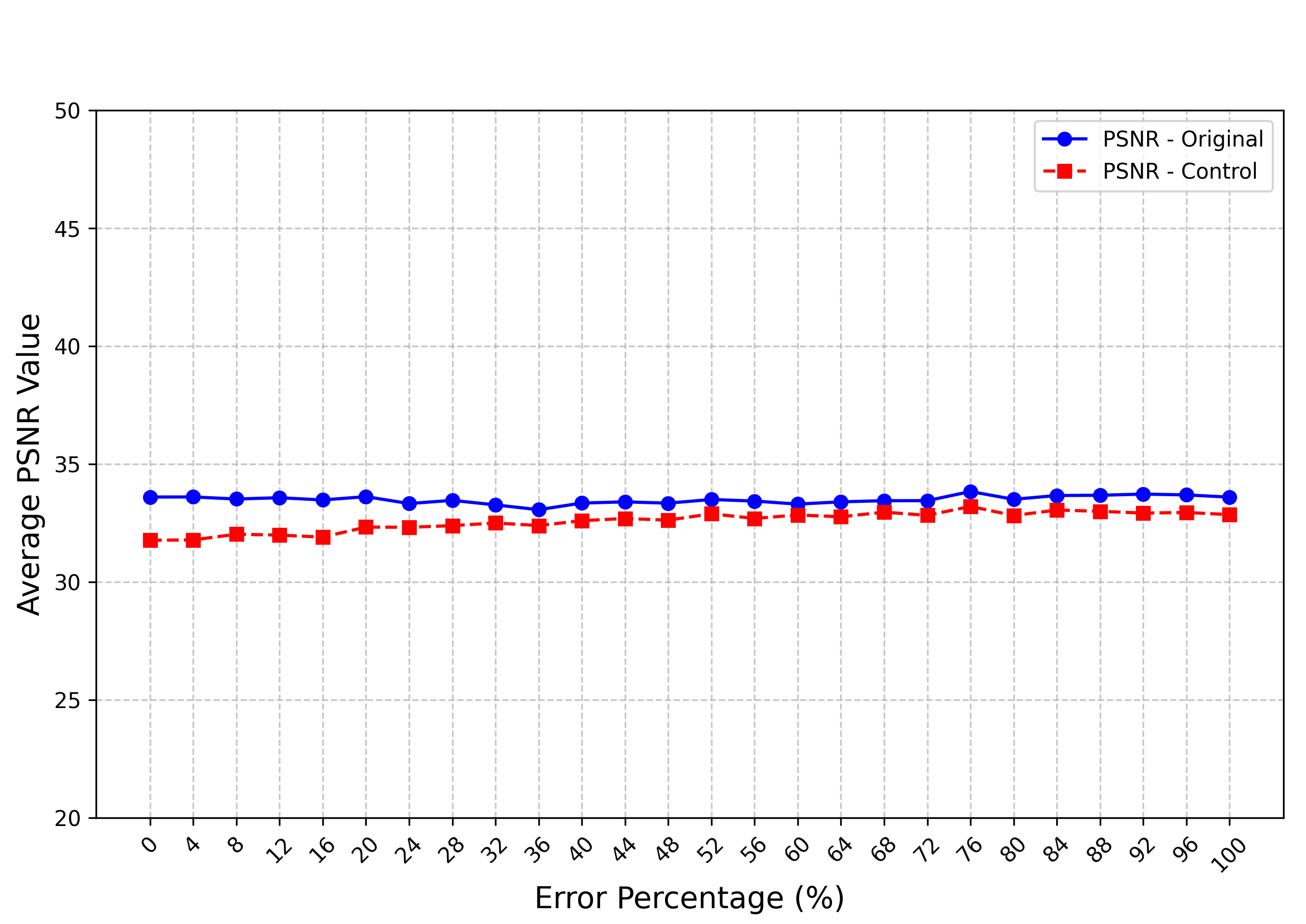}
\centering
\caption{Average PSNR vs Error Percentage (Error Type 1).}
\captionsetup{justification=centering}  
\label{fig:PSNR_et3}
\end{figure}

\begin{figure}[h!]
\includegraphics[width=0.47\textwidth]{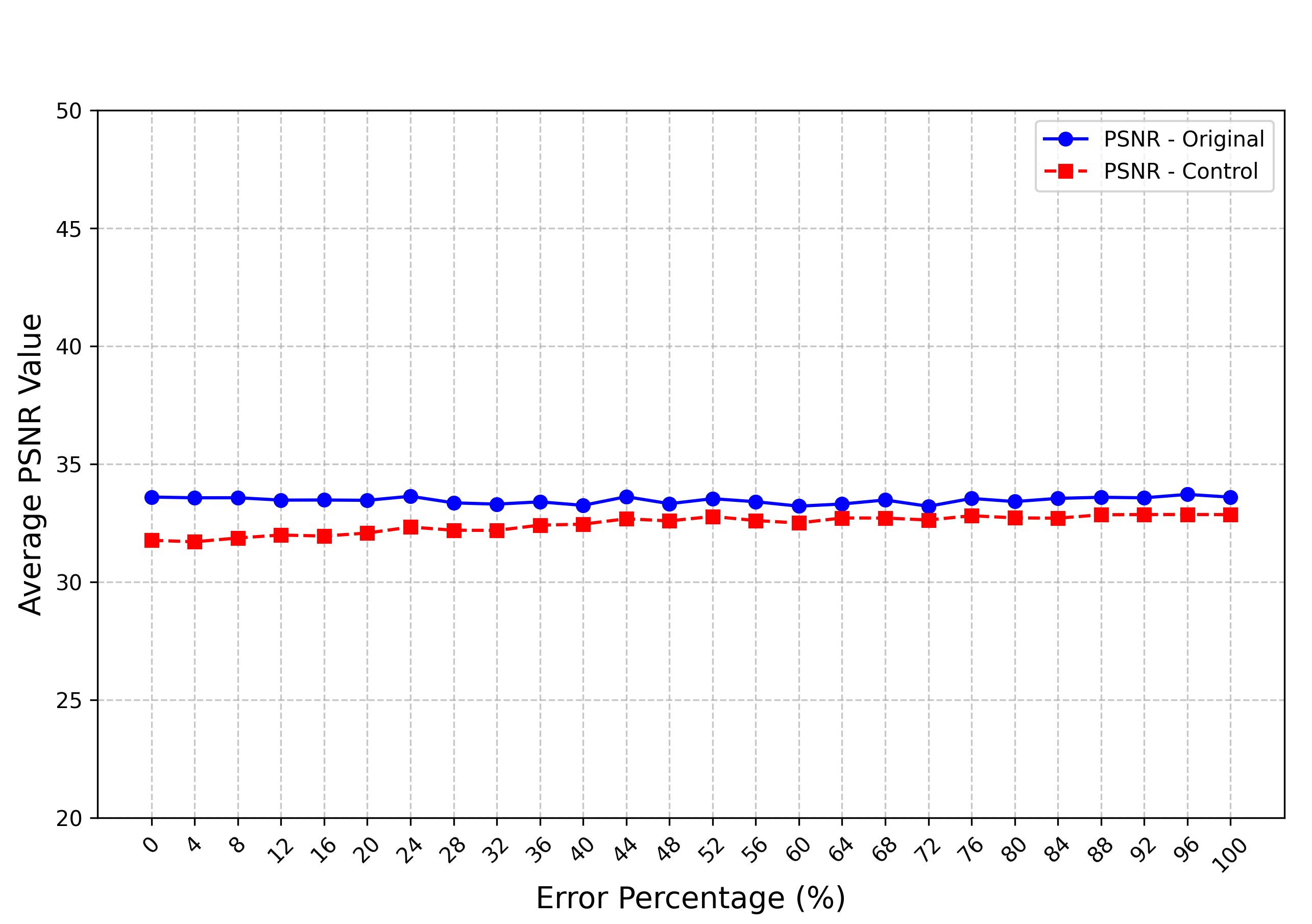}
\centering
\caption{Average PSNR vs Error Percentage (Error Type 2).}
\captionsetup{justification=centering}  
\label{fig:PSNR_et4}
\end{figure}

\begin{figure}[h!]
\includegraphics[width=0.47\textwidth]{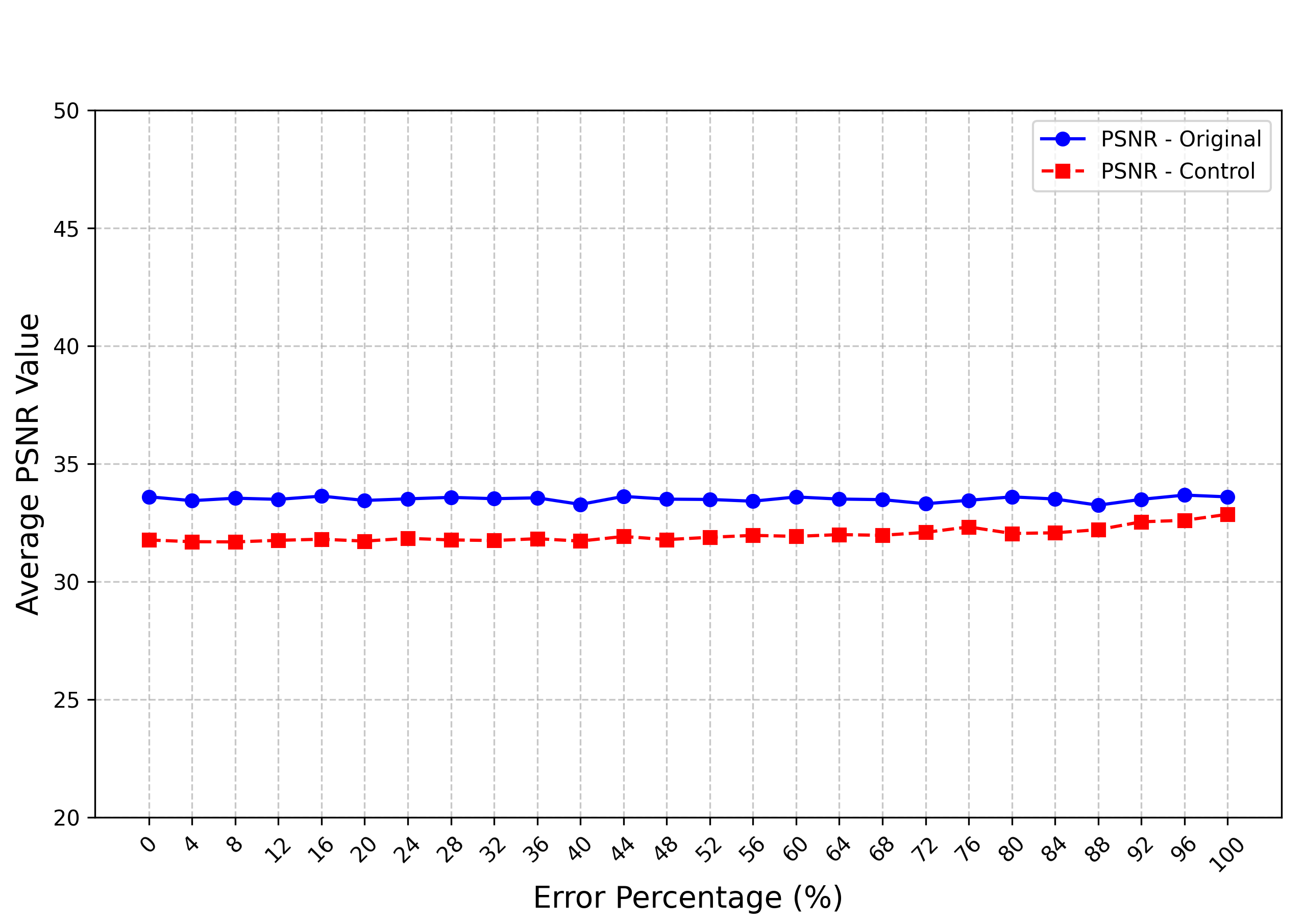}
\centering
\caption{Average PSNR vs Error Percentage (Error Type 3).}
\captionsetup{justification=centering}  
\label{PSNR_et6}
\end{figure}


\begin{figure}[h!]
\includegraphics[width=0.47\textwidth]{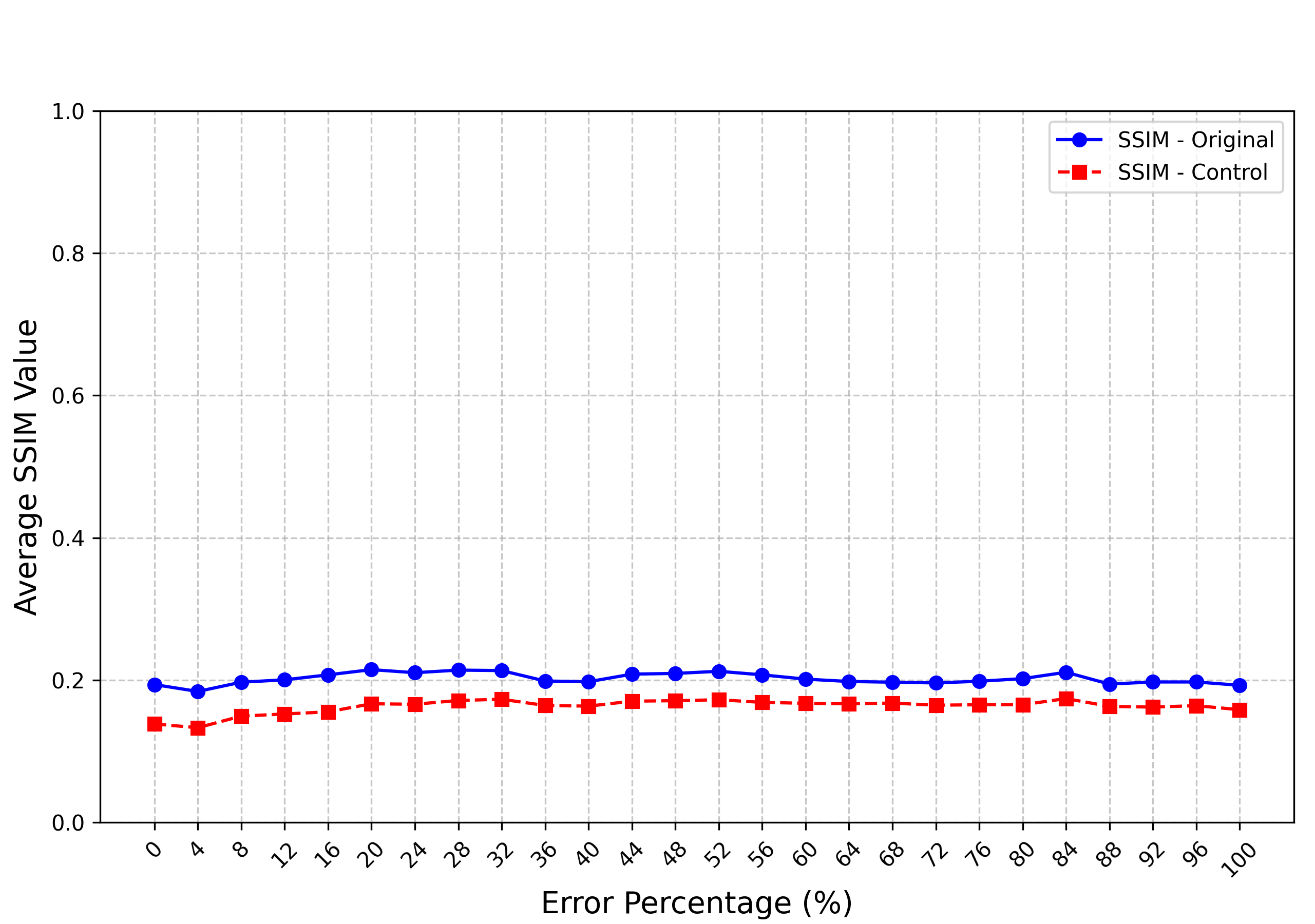}
\centering
\caption{Average SSIM vs Error Percentage (Error Type 1).}
\captionsetup{justification=centering}  
\label{SSIM_et3}
\end{figure}

\begin{figure}[h!]
\includegraphics[width=0.47\textwidth]{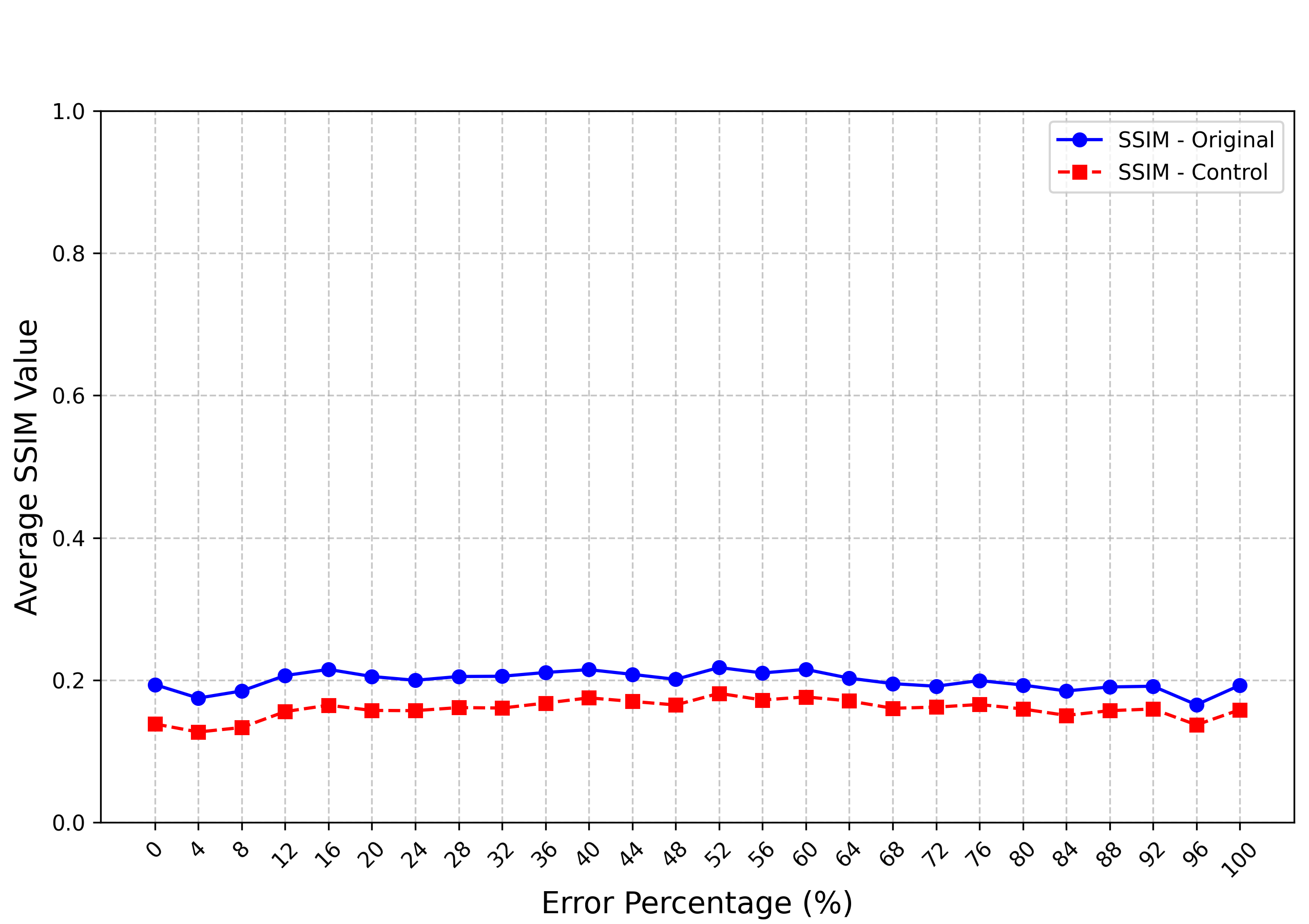}
\centering
\caption{Average SSIM vs Error Percentage (Error Type 2).}
\captionsetup{justification=centering}  
\label{SSIM_et4}
\end{figure}

\begin{figure}[h!]
\includegraphics[width=0.47\textwidth]{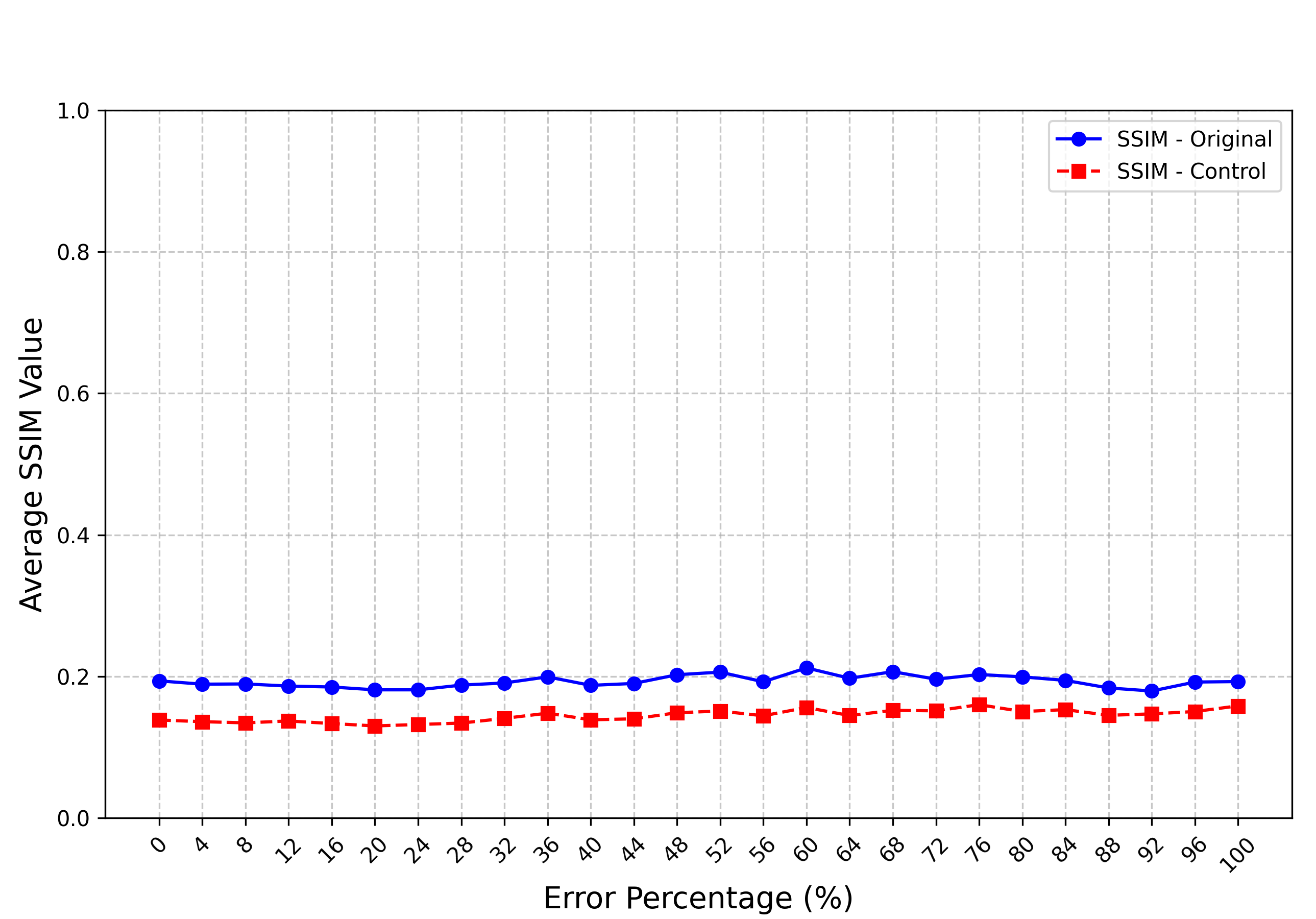}
\centering
\caption{Average SSIM vs Error Percentage (Error Type 3).}
\captionsetup{justification=centering}  
\label{SSIM_et6}
\end{figure}

\begin{figure}[h!]
\includegraphics[width=0.47\textwidth]{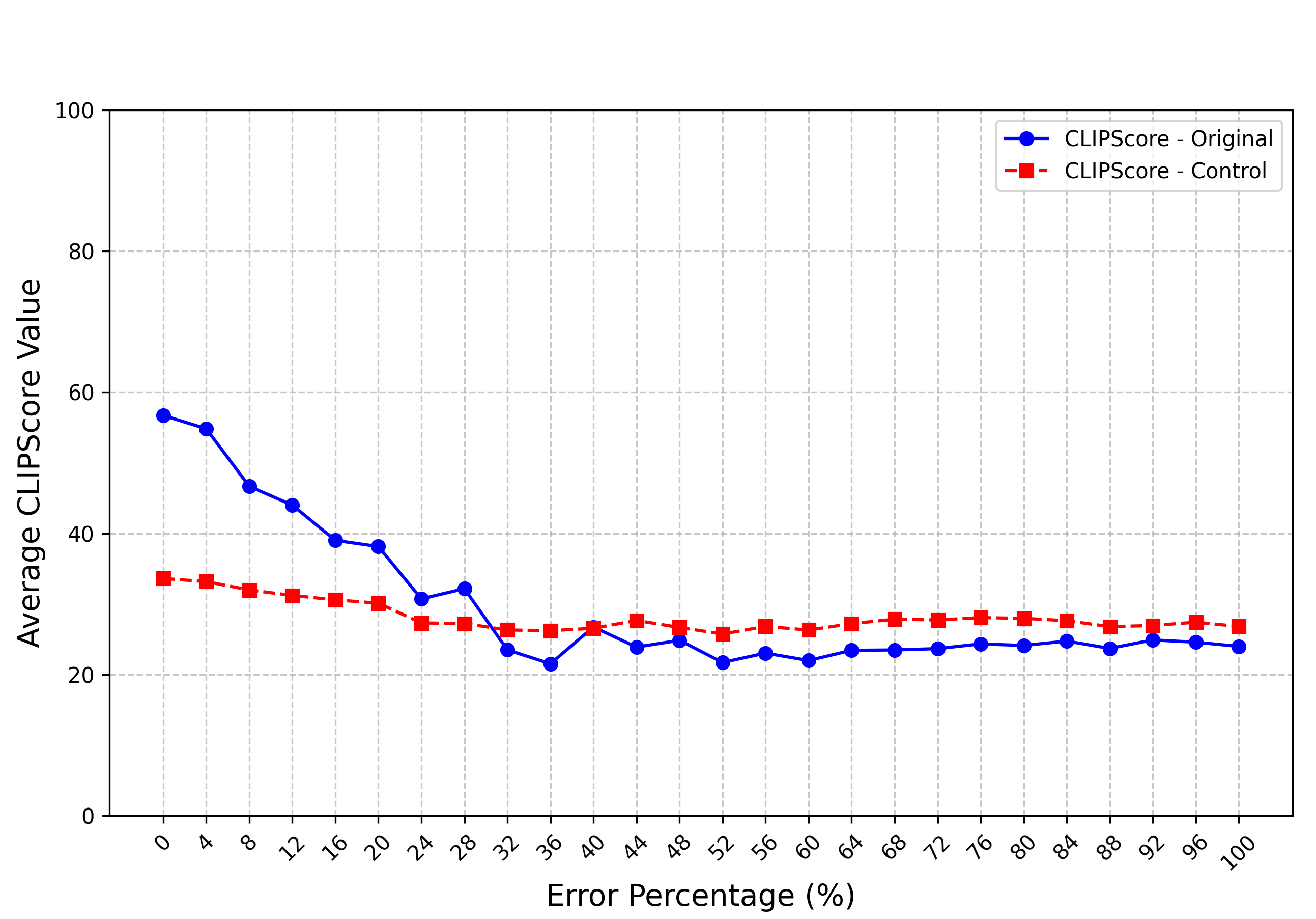}
\centering
\caption{Average CLIP Similarity Score vs Error Percentage (Error Type 1).}
\captionsetup{justification=centering}  
\label{CLIPScore_et3}
\end{figure}

\begin{figure}[h!]
\includegraphics[width=0.47\textwidth]{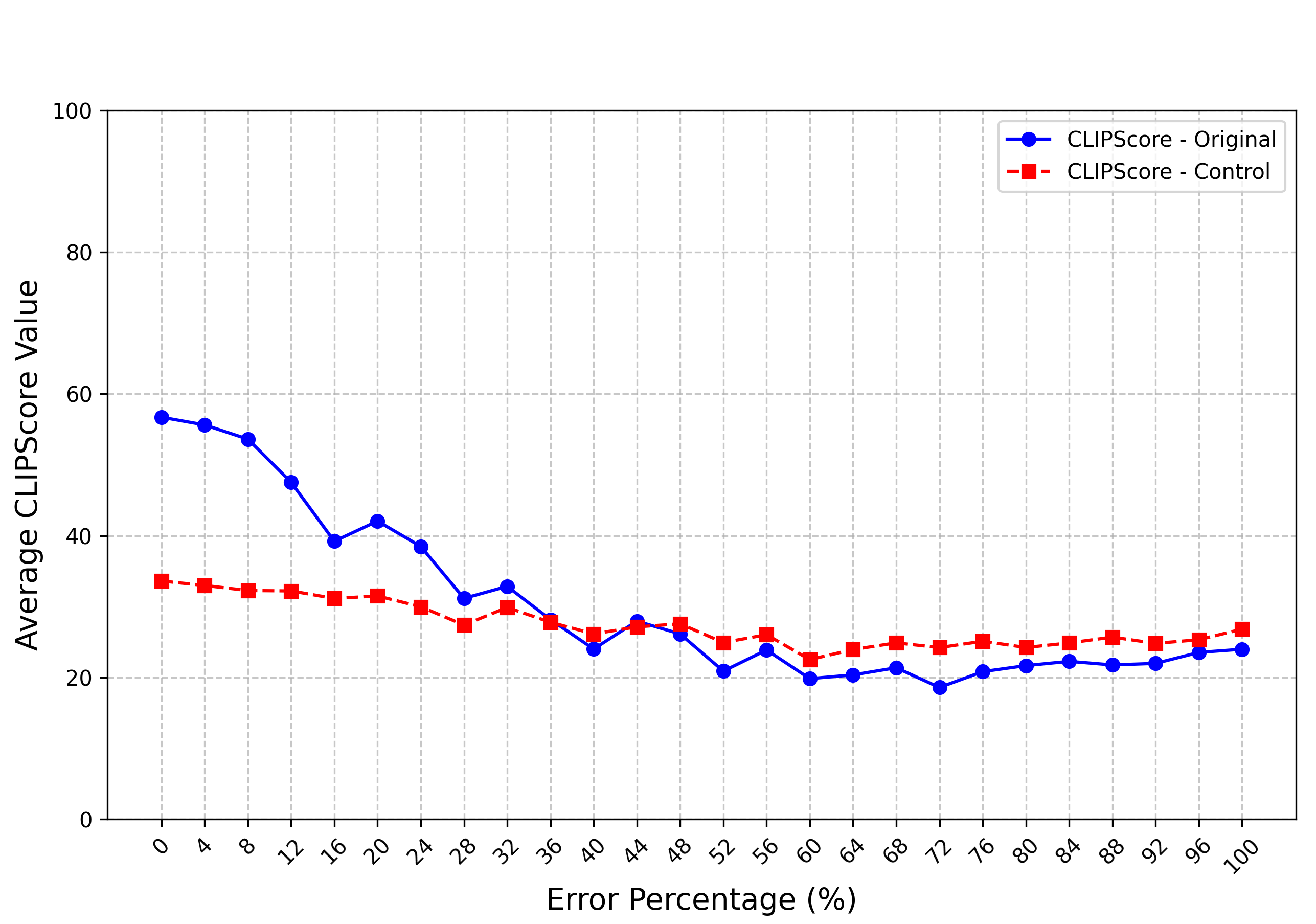}
\centering
\caption{Average CLIP Similarity Score vs Error Percentage (Error Type 2).}
\captionsetup{justification=centering}  
\label{CLIPScore_et4}
\end{figure}

\begin{figure}[h!]
\includegraphics[width=0.47\textwidth]{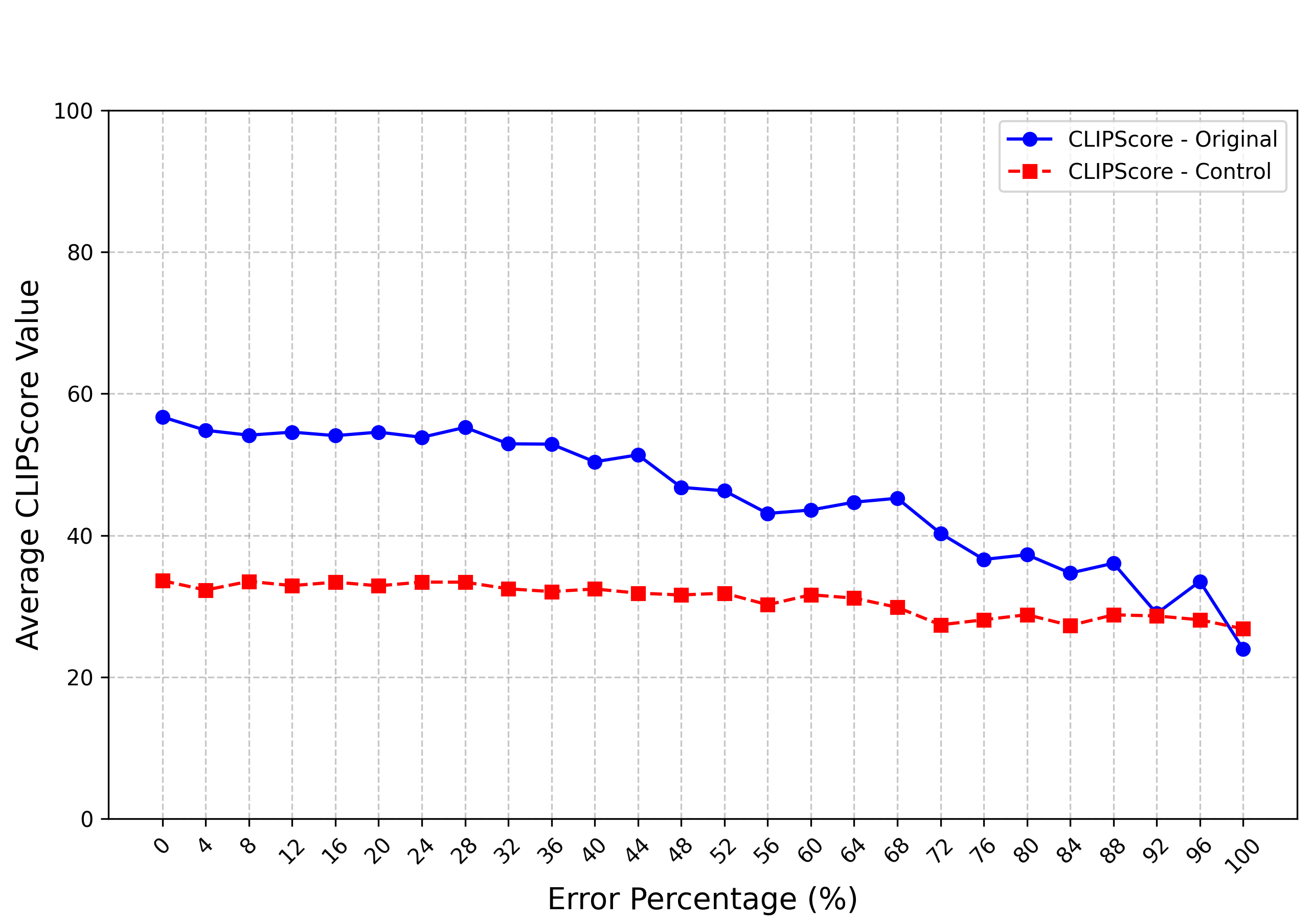}
\centering
\caption{Average CLIP Score vs Error Percentage (Error Type 3).}
\captionsetup{justification=centering}  
\label{CLIPScore_et6}
\end{figure}

\subsection{Evaluation Analysis}

Our analysis reveals that both PSNR and SSIM exhibit minimal sensitivity to semantic degradation (see Figs. \ref{fig:PSNR_et3} to \ref{SSIM_et6}). Although suitable for measuring traditional image compression quality, they are inadequate for comparing semantically similar but visually different images. This limitation becomes evident in Fig.~\ref{tabela_exemplos} and aligns with the findings discussed in our previous work~\cite{sage}. 

CLIPScore better reflects perceptual similarity and semantic consistency. For example, in Fig. \ref{CLIPScore_et3}, which shows results under random character substitution (Error Type 1), CLIPScore decreases from 60\% to 50\% when the CER reaches approximately 15\%. In Fig.~\ref{CLIPScore_et4}, representing character deletion (Error Type 2), a similar drop occurs after CER reaches 20\%. Finally, in Fig.~\ref{CLIPScore_et6}, where word deletion (Error Type 3) is simulated, the same degradation only appears when the CER approaches 50\%.

Our findings suggest that character-level corruption has a more significant impact on semantic understanding compared to word-level deletions. This is likely because not all words contribute equally to the semantic meaning -- removal of non-essential words (e.g., articles or conjunctions) has a smaller impact than partial corruption of multiple words. 

\subsection{Key Findings}
The obtained results demonstrate that SAGE can tolerate a substantial degree of textual corruption while maintaining meaningful semantic output. In particular, CLIPScore results confirm that semantic image reconstruction remains viable up to moderate error ratios (e.g., 15–20\% CER), reinforcing the potential of semantic communication frameworks like SAGE in noisy underwater acoustic environments.

\section{Conclusions}
This paper highlights the resilience of the SAGE framework for underwater wireless communications, where conventional transmission methods face severe limitations due to harsh propagation conditions. By combining semantic processing and Generative AI, SAGE enables the compression of visual data into compact textual descriptions and robustly reconstructs image content even under significant transmission errors. 

Our evaluation results confirm that SAGE can withstand substantial degradation while still maintaining meaningful image reconstruction. This resilience reduces the need for heavy error detection and correction mechanisms. These findings demonstrate the potential of semantic-based approaches to enhance communication reliability and efficiency in challenging underwater environments. Future work will focus on enhancing error resilience even further, exploring advanced correction techniques and refining the generative models for better performance.

\bibliographystyle{IEEEtran}
\bibliography{myrefs}

\end{document}